# Anomalous sign inversion of spin-orbit torque in ferromagnetic/nonmagnetic bilayer systems due to self-induced spin-orbit torque

Motomi Aoki[1], Ei Shigematsu[1], Ryo Ohshima[1], Teruya Shinjo[1], Masashi Shiraishi[1], and Yuichiro Ando[1,2†]

[1]Department of Electronic Science and Engineering, Kyoto University, Kyoto, Kyoto 615-8510, Japan

[2]PRESTO, Japan Science and Technology Agency, Honcho, Kawaguchi, Saitama 332-0012, Japan

**Corresponding authors**

[†]**Yuichiro Ando** Address: A1-226, Kyodai Katsura, Nishikyo-ku, Kyoto, Kyoto 615-8510, Japan

Tel.: +81-75-383-2356, Fax: +81-75-383-2275

E-mail: ando.yuichiro.5s@kyoto-u.ac.jp

Self-induced spin-orbit torques (SI-SOTs) in ferromagnetic (FM) layers have been overlooked when estimating the spin Hall angle (SHA) of adjacent nonmagnetic (NM) layers. In this work, we observe anomalous sign inversion of the total SOT in the spin-torque ferromagnetic resonance due to the enhanced SI-SOT, and successfully rationalize the sign inversion through a theoretical calculation considering the SHE in both the NM and FM layers. The findings show that using an FM layer whose SHA sign is the same as that of the NM achieves efficient SOT-magnetization switching with the assistance of the SI-SOT. The contribution of the SI-SOT becomes salient for a weakly conductive NM layer, and conventional analyses that do not consider the SI-SOT can overestimate the SHA of the NM layer by a factor of more than 150.

# I. Introduction

The spin Hall effect (SHE) [1] in a nonmagnetic (NM) material with a sizable spin-orbit interaction (SOI) realizes injection of a pure spin current into an adjacent ferromagnetic (FM) material and exerts torque on the magnetization via spin-orbit torque (SOT) [2], which enables manipulation and even switching of the magnetization [3–10]. Efficient generation of SOT requires a material with a large spin Hall angle (SHA). Because the magnitude of SOI is roughly proportional to the fourth power of the atomic number, most research on SOT has focused on materials containing heavy elements [5,7,11,12]. Highly efficient charge-to-spin conversion has been discovered in platinum(Pt), tungsten(W), and tantalum (Ta) [5,6,11]. Such studies have generally used FM/NM bilayer systems to directly detect the SOT. However, most of these studies have overlooked a non-negligible contribution: the self-induced spin-orbit torque (SI-SOT), which originates from the spin Hall effect in the FM layer itself. Furthermore, recent studies have revealed that charge-to-spin conversion efficiency in 3d FM layers is substantially high despite their relatively small atomic numbers [13–16], and the SHE of the FM layer can exert SOT on the FM magnetization itself in an NM/FM bilayer [17–19]. Nevertheless, no experimental studies have addressed non-negligible SI-SOT in the FM layer and SOT from the adjacent NM layer separately. This hampers precise estimation of the SHA of the NM layer, because SOT applied to the FM layer is a combination of the aforementioned SOTs with different physical origins.

In this paper, we demonstrate anomalous sign inversion of the SOT in NM/FM bilayer devices in spin-torque ferromagnetic resonance (ST-FMR) [11], which is caused by enhanced SI-SOT. Since the SI-SOT (scaled by the spin-dephasing length in the FM) and SOT from the NM layer (scaled by the spin-diffusion length in the NM) exhibit different thickness dependences, these contributions can be separated by measuring the ST-FMR signals with a wide range of thickness of the FM layer, $t_{FM}$. The $t_{FM}$ dependence of the total SOT was well reproduced by a theoretical model considering the SHE in both the NM and FM layers [20]. Consequently, the SI-SOTs are –43 and 27% of the SOT from the NM layer in Ta(5 nm)/Co(5 nm) and permalloy(5 nm)/Pt(5 nm) bilayers, respectively, which are both large. Our findings reveal that combining FM and NM layers with the same SHA sign achieves efficient SOT-magnetization switching, because the SI-SOT augments the efficiency. More importantly, the SI-SOT contribution becomes dominant for a weakly conductive NM layer, and the conventional analyses of the SOT in NM/FM bilayer systems that do not consider SI-SOT can overestimate the SHA of the NM layer by a factor of more than 150. Our study provides a fuller understanding of conventional SOT physics.

**II. Experimental procedures**

Figure 1(a) shows a schematic of the device structure and the electrical circuit used in our study. Rectangular 10 μm × 25 μm Ta($t_{NM}$)/Co($t_{FM}$)/SiO$_2$(7 nm) channels were fabricated on MgO (001) substrates using rf-magnetron sputtering, where $t_{NM}$ was fixed to 5 nm and $t_{FM}$ was varied from 3 to 17.5 nm. Hereafter, number in bracket indicates thickness in the unit of nanometers. In the ST-FMR experiments, a DC voltage, $V_{DC}$, was measured under an introduced microwave AC current using a commercial analog signal generator via a Ti(3)/Au(70) coplanar wave guide. The angle $\theta$, between the external magnetic field $H_{ext}$ and the $x$ axis, was changed from 0° to 360°. All measurements were carried out at room temperature. $V_{DC}$ is expressed as [11]

$$V_{DC} = A \frac{\Delta(\mu_0 H_{ext} - \mu_0 H_{FMR})}{(\mu_0 H_{ext} - \mu_0 H_{FMR})^2 + \Delta^2} + S \frac{\Delta^2}{(\mu_0 H_{ext} - \mu_0 H_{FMR})^2 + \Delta^2}, \qquad (1)$$

where $A$ and $S$ are the magnitudes of the anti-symmetric and the symmetric Lorentzian functions, respectively, $\Delta$ is the half-width at half maximum, $\mu_0$ is the vacuum permeability, and $H_{FMR}$ is the ferromagnetic resonance field. Here, we use the definition of FMR spin-torque efficiency, $\xi_{FMR}$, which has been considered to be close to the SHA of the NM layer, as [11]

$$\xi_{FMR} = \frac{S}{A} \frac{e\mu_0 M_S t_{NM} t_{FM}}{\hbar} \sqrt{1 + \frac{M_{eff}}{H_{FMR}}}, \qquad (2)$$

where $e$, $M_S$, $M_{eff}$, and $\hbar$ are the elementary charge, saturation magnetization, effective magnetization, and Dirac constant, respectively. $M_{eff}$ was obtained from the $H_{FMR}$–$f$ curve using the Kittel formula [21], and $\mu_0 M_S$ for Co was determined to be 1.87 T from the linear fit of the $1/M_{eff}$–$1/t_{FM}$ plot [22] (see Supplemental Material (SM) Sec. A [23]).

**III. Spin-torque ferromagnetic resonance measurement on Ta/Co bilayers**

Figure 1(b) shows the ST-FMR spectra for $t_{FM}$ = 3 (left), 5 (middle), and 17.5 nm (right) when $\theta$ = 45°. The red (blue) curves are the symmetric (antisymmetric) components of the spectra obtained by fitting with Eq. (1). When the SI-SOT is negligible, the sign of $A(S)$ is positive (negative) in our setup, considering the direction of the Oersted field and the negative SHA of Ta. The signs of both $A$ and $S$ in Ta(5)/Co(5) agree with the expectation. In Ta(5)/Co(3), the sign of $A$ was inverted owing to the contribution from the fieldlike SOT [24], but the sign of $S$ was negative as in Ta(5)/Co(5). A significant result is in the ST-FMR spectrum of Ta(5)/Co(17.5), where the sign of $S$ was surprisingly inverted by simply increasing the Co thickness [see the right panel of Fig. 1(b)]. To reveal $t_{FM}$ dependence of the spin torque in detail, $\theta$ dependences of ST-FMR signals were measured. Table I summarizes possible angular dependences in antisymmetric ($A$) and symmetric (*S*) component in the ST-FMR spectrum, where ***M*** is the magnetization vector [25,26]. Under the experimental setup shown in Fig. 1(a), Oersted field along the *y* direction and SOT from $\sigma_y$ spins via the SHE in Ta and Co are expected to be dominant contributions. Both Oersted field along the *y* direction and fieldlike (FL) SOT due to $\sigma_y$ spins contribute to $\sin2\theta\cos\theta$ term in *A*, while dampinglike (DL) SOT due to $\sigma_y$ spins contribute to $\sin2\theta\cos\theta$ term in *S*. Though other contributions such as Oersted field along the *x* and *z* directions are also expected, they do not appear in $\sin2\theta\cos\theta$ term of *S*. Because the inverse spin Hall effect induced by spin pumping (SP-ISHE) and anomalous Nernst effect (ANE) are negligible in Ta/Co bilayers used in our study as discussed later, $\xi_{FMR}$ is estimated more precisely by using $A(S)_{\sin2\theta\cos\theta}$ as $A(S)$ in Eq. (2), where $A(S)_{\sin2\theta\cos\theta}$ is $\sin2\theta\cos\theta$ term in $A(S)$. Figures 1(c) and 1(d) show $\theta$ dependence of *A*, while the upper panels in Figs. 1(e) and 1(f) show $\theta$ dependences of *S* for Ta(5)/Co(5) and Ta(5)/Co(17.5), respectively, where solid lines are total fitting (red), $\sin2\theta\cos\theta$ (blue), $\sin2\theta$ (green), $\sin2\theta\sin\theta$ (purple), and $\sin\theta$ (brown) term (see Table I). Schematics of expected $\sin2\theta\cos\theta$ component in *S* with negative and positive SOTs are also exhibited in the lower panels of Figs. 1(e) and 1(f), respectively. Whereas the sign of $A_{\sin2\theta\cos\theta}$ was identical in both cases, the sign of $S_{\sin2\theta\cos\theta}$ was actually inverted for $t_{FM}$ = 17.5 nm, which cannot be explained in the conventional ST-FMR framework.

To explain the anomalous sign inversion in $\sin2\theta\cos\theta$ component in *S*, shown in Figs. 1(e) and 1(f), contribution of the spin Hall effect in the Co layer is considered. Figure 1(g) shows a schematic of spin-current generation via the SHE of Ta with a negative SHA in the Ta/Co bilayer. The electric current in the Ta layer along the +*x* direction, $J_{c(Ta)}$, is converted into the spin current along the –*z* direction, $J_{s(Ta)}$, via the SHE in the Ta layer, resulting in the negative SOT. Here, negative (positive) SOT is defined to be the SOT that aligns the magnetization along the +(–)*y* direction. Note that the electric shunting current through the Co layer is a function of the conductance ratio between the Co and Ta layers, yielding spin current via the SHE of Co. The electric current shunting into the Co layer along the +*x* direction, $J_{c(Co)}$, is converted into the spin current along the +*z* direction, $J_{s(Co)}$ [see Fig. 1(h)]. In the case of a single Co layer without adjacent NM layers, spins with opposite directions (+*y* and –*y*) accumulate on the top and the bottom surfaces, and they cannot flow out from the Co layer. Consequently, the net sum of the spin accumulations in the Co layer is zero. In contrast, in the case of the Ta/Co bilayer, spins accumulated at the Ta/Co interface can diffuse into the Ta layer through the Ta/Co interface, and a net spin accumulate at the top interface, generating SI-SOT with positive polarity. The amount of the SOT of the NM layer and the SI-SOT depends on $t_{NM}/l_{NM}$, and the $t_{FM}/l_{FM}$, respectively, as well as the SHA of the NM and FM layer, where $l_{NM(FM)}$ is the spin-diffusion (dephasing) length of the NM (FM). Thus, the experimentally detected SOT, i.e., the net SOT, can be controlled by changing $t_{FM}$. Increasing $t_{FM}$ allows the cancellation of the SOT of the NM layer by the SI-SOT, and the anomalous sign inversion in $\sin2\theta\cos\theta$ component in *S*.

## IV. Theoretical analysis of thickness dependence of $\xi_{FMR}$ using the spin-diffusion model

Figure 2(a) shows $\xi_{FMR}$ obtained from $\theta$ dependence of the ST-FMR spectra as a function of $t_{FM}$ for the various Ta/Co bilayers. $\xi_{FMR}$ at $t_{FM}$ = 3 nm is positive because of negative $A$, which is attributed to the fieldlike SOT [24]. Above $t_{FM}$ = 5 nm, $\xi_{FMR}$ is negative because of positive $A$ due to dominant contribution of the Oersted field, and negative $S$. Importantly, $\xi_{FMR}$ becomes positive again at $t_{FM} \geq$ 8.5 nm owing to the sign reversal of $S$, which is attributed to the SI-SOT. Because both the DL torque efficiency, $\xi_{DL}$, and FL torque efficiency, $\xi_{FL}$, are known to have negligible dependence on $t_{FM}$ in the conventional understanding, $1/\xi_{FMR}$ is expected to be linearly proportional to $1/t_{FM}$. This has been used to estimate $\xi_{DL}$ and $\xi_{FL}$ from the linear fit of the $1/\xi_{FMR}$–$1/t_{FM}$ plot [24,27,28]. However, the results in Fig. 2(b) show a noticeable nonlinear dependence of $1/\xi_{FMR}$ on $1/t_{FM}$. Thus, we calculated $\xi_{FMR}$ using the spin-diffusion equation [20] applying the following boundary conditions: continuity of the spin chemical potential and spin current at the NM/FM interface, and zero spin current at the top and bottom of the NM/FM bilayer. The total spin current with a spin vector transverse to the magnetization at the NM/FM interface, $J_{s\perp}$, is expressed as

$$J_{s\perp} = \frac{\tanh(\frac{t_{NM}}{2l_{NM}})R_{s(NM)}J_{c(NM)}\theta_{NM} + \tanh(\frac{t_{FM}}{2l_{FM}})R_{s(FM)}J_{c(FM)}\theta_{FM}}{R_{s(NM)}\coth\left(\frac{t_{NM}}{l_{NM}}\right) + R_{s(FM)}\coth\left(\frac{t_{FM}}{l_{FM}}\right)}\sin\theta. \quad (3)$$

Here, $\theta_{NM}$ ($\theta_{FM}$) is the SHA of the NM (FM) layer, and $R_{s(NM)}$ ($R_{s(FM)}$) $\equiv l_{NM(FM)}/\sigma_{NM(FM)}$ is the spin resistance of the NM (FM) layer, where $\sigma_{NM(FM)}$ is the conductivity of the NM (FM) layer. For simplicity, we neglect the interfacial spin-orbit coupling and the spin precession during the diffusion [29], i.e., $\xi_{DL} = J_{s\perp}/(J_{c(Ta)}\sin\theta)$, and thus, Eq. (3) is the simplified expression of the generalized formalism described in Ref. [18]. Such a simplification does not affect our main results because the contribution of the interfacial spin-orbit coupling is much less dependent on NM and FM thicknesses. Consequently, $\xi_{FMR}$ is expressed as [27]

$$\xi_{FMR} = \left\{\frac{R_{s(NM)}\coth\left(\frac{t_{NM}}{l_{NM}}\right) + R_{s(FM)}\coth\left(\frac{t_{FM}}{l_{FM}}\right)}{\tanh\left(\frac{t_{NM}}{2l_{NM}}\right)R_{s(NM)}\theta_{NM} + \tanh\left(\frac{t_{FM}}{2l_{FM}}\right)R_{s(FM)}\frac{\sigma_{FM}}{\sigma_{NM}}\theta_{FM}} + \frac{\hbar}{e\mu_0 M_s t_{NM} t_{FM}}\frac{\xi_{FL}}{\xi_{DL}}\right\}^{-1}. \quad (4)$$

Figures 2(c) and 2(d) show $\xi_{FMR}$ calculated as a function of $t_{FM}$ for Ta/Co for (c) various values of $\theta_{FM}$ (0.01, 0.05, and 0.1) and fixed $l_{FM}$ (3 nm) (case 1A), and (d) fixed $\theta_{FM}$ (0.05) and various $l_{FM}$ (2, 3, and 5 nm) (case 1B) [30,31]. Here, we used literature values of $l_{NM}$ and $\theta_{NM}$ [5,32], and the measured values for $\sigma_{NM}$ and $\sigma_{FM}$ (case 1 in Table II). Although $\theta_{FM}$ measured using the spin valve is on the order of 0.01 [15], it is on the order of 0.1 from the measurement of the anomalous spin-orbit torque [31] and the theoretical calculation [33]. Therefore, $\theta_{FM}$ should be varied from 0.01 to 0.1. Considering that the reported values of $\xi_{FL}/\xi_{DL}$ are from –1 to 1 [24,27] and the fieldlike SOT is large enough to cancel the Oersted field when $t_{FM}$ < 3 nm, we set $\xi_{FL}/\xi_{DL}$ = 1 for the Ta/Co. Both cases 1A and 1B qualitatively reproduce the experimental results obtained from ST-FMR as shown in Fig. 2(a): the sign reversal of $\xi_{FMR}$ appearing in the thin and thick regimes is related to the fieldlike SOT and the SI-SOT, respectively. Using $\theta_{FM}$ = 0.05, $\theta_{NM}$ = –0.15, $l_{FM}$ = 3 nm, and $l_{NM}$ = 1.8 nm, we find the SI-SOT contribution to $\xi_{DL}$ in the Ta(5)/Co(5) sample is approximately –43% (the negative sign indicates suppression of the SOT from Ta) of that of the SOT arising from Ta, indicating that the SI-SOT largely hampers the SOT from the Ta layer. The presence or absence of the sign reversal of $\xi_{FMR}$ from negative to positive strongly depends on the values of $\theta_{FM}$, $\theta_{NM}$, $l_{FM}$ and $l_{NM}$. The conditions for the sign reversal are large $\theta_{FM}$ and $l_{FM}$, and small $\theta_{NM}$ and $l_{NM}$, because larger $\theta_{FM}$ results in more efficient spin-current generation in the FM layer and larger $l_{FM}$ is equivalent to a higher spin resistance of the FM layer. This enhances the spin-current flow from the FM to the NM layer. The small $\theta_{NM}$ and $l_{NM}$ also contribute to the SOT suppression and low spin resistance of the NM layer, respectively.

## V. Control experiments and contribution of spurious effects

To obtain further evidence, we investigated the thickness dependence of $\xi_{FMR}$ for permalloy (Ni$_{81}$Fe$_{19}$, Py)/Pt bilayers because no anomalous sign inversion is expected owing to positive $\theta_{NM}$ and $\theta_{FM}$. Figures 3(a) and 3(b) show $\xi_{FMR}$ and $1/\xi_{FMR}$ as functions of $t_{FM}$ and $1/t_{FM}$, respectively, for Py ($t_{FM}$)/Pt (5 nm). Sign inversion of $\xi_{FMR}$ indeed was not observed up to $t_{FM}$ = 15 nm. Meanwhile, the $1/\xi_{FMR}$–$1/t_{FM}$ plot does not exhibit a linear relationship as shown in Fig. 3(b). Calculating $\xi_{FMR}$ using Eq. (4) well reproduces the experimental result even for thin and thick $t_{FM}$ regions [see Figs. 3(c) and 3(d) and Table II (case 2) for the calculation parameters]. Using $\theta_{FM}$ = 0.05, $\theta_{NM}$ = 0.32, $l_{FM}$ = 5 nm, and $l_{NM}$ = 1.4 nm, we estimate the contribution of SI-SOT to $\xi_{DL}$ in the Py(5)/Pt(5) sample to be +27% of that of the SOT from Pt, indicating that the SI-SOT assists the original SOT from the Pt layer. Importantly, the enhancement of $\xi_{FMR}$ for larger $\theta_{FM}$ in Fig. 3(c) shows that selecting an FM layer with a large $\theta_{FM}$ that has the same sign as $\theta_{NM}$ achieves efficient SOT generation with assistance from the SI-SOT.

Other possible origins of the observed sign inversion are SP-ISHE [34], ANE [35], the unidirectional spin Hall magnetoresistance (USMR) [36], and the orbital Hall effect (OHE) [37,38]. These effects reportedly become pronounced at large $t_{FM}$ [39–41]. However, they are discernible from the SI-SOT studied in this work as follows. From the value of spin-mixing conductance between Ta/Co interface, $1.7\times10^{-19}$ m$^{-2}$, which was estimated from $t_{FM}$ and $t_{NM}$ dependences of the damping coefficient, SP-ISHE was calculated to be less than 15 % of $S$ component in our thickness range, which cannot explain the sign inversion (see SM Secs. A and B [23]) [42–45]. In addition, the sign of $S$ via the ANE and the USMR should be negative [45], which does not rationalize the sign inversion in our Ta/Co. The OHE and resulting torque might also contribute to this phenomenon [37,38]. However, the orbital current injected into the FM layer is rapidly converted into spin current within a thickness of a few atomic layers [37]. In such a case, although the net of spin current injected into the FM layer might be modulated owing to the OHE contribution, its effect is independent of $t_{FM}$ when $t_{FM}$ is thicker than the orbital diffusion length, typically a few atomic layers. This means that the OHE is not responsible for the sign inversion when $t_{FM} > 10$ nm. Therefore, we conclude that the aforementioned effects are negligible.

We also verified the influence of the SI-SOT from the shift of the magnetoresistance curve [46] and the second harmonic Hall measurements [47,48] (see SM Secs. C and D [23]). In addition, we confirmed that sign of the magnetoresistance and/or crystalline structure are unchanged even when $t_{FM}$ was increased up to 17.5 nm (see SM Secs. E and F [23]). Furthermore, we observed sign inversion of $\xi_{FMR}$ due to the SI-SOT in Pt/Fe bilayer (see SM Sec. G [23]). These control experiments also support our claim that the sign inversion of $S$ in the thick FM is attributed to the enhancement of the SI-SOT. We note that even when the SOT in a single FM layer [31,49–51] is negligibly small, as is previously reported and actually obtained in our single Co layer (see SM section H [23]), the contribution of the SI-SOT is not negligible because the spin-current absorption to the adjacent NM layer is essential for generation of the SI-SOT.

## VI. Overestimation of spin Hall angle of nonmagnetic materials due to contribution of self-induced spin-orbit torques

Finally, we generalize the discussion of $\xi_{FMR}$ to a wide variety of bilayer systems investigated in recent SOT studies. The nonlinear relationships between $1/\xi_{FMR}$ and $1/t_{FM}$, in principle, negate the validity of the conventional estimation of $\xi_{FL}$ from linear fitting of the $1/\xi_{FMR}$–$1/t_{FM}$ curve because the SI-SOT is non-negligible. We also emphasize that observing a linear relationship in the $1/\xi_{FMR}$–$1/t_{FM}$ curve does not always mean the absence of SI-SOI, and reliability of estimating $\xi_{FL}$ from linear fitting is a little subtle. To discuss this uncertainty, we calculate $1/\xi_{FMR}$–$1/t_{FM}$ curves for different values of $\xi_{FL}/\xi_{DL}$ in Fig. 4(a), where $1/\xi_{FMR}$ has almost a linear relation to $1/t_{FM}$ even for $\xi_{FL}/\xi_{DL} = 0$. In this case, the value of $\xi_{FL}$ estimated via conventional analysis is non-zero but significantly deviates from the real value, $\xi_{FL} = 0$, unless SI-SOT is considered. Further model calculation reveals substantial overestimation of the SHA when the SI-SOT is neglected. case 4A shown in Fig. 4(b) is an example of a topological insulator (TI) [7,8] where $\theta_{NM}$ is changed from 0.001 to 0.1 and $\theta_{FM}$ and $\sigma_{NM}$ are fixed to 0.1 and $1.76 \times 10^4$ ($\Omega \cdot$m)$^{-1}$, respectively, i.e., the weakly conductive NM regime. Importantly, large $\xi_{FMR}$ up to 0.15 was obtained even for $\theta_{NM} = 0.001$. In the conventional analysis, $\xi_{FMR} \sim \theta_{NM}$ has been postulated so far for $\xi_{FL} \sim 0$. Therefore, the calculated result in Fig. 4(b) indicates that $\theta_{NM}$ is overestimated by a factor of about 150 if the SI-SOT is neglected. The overestimation is due to neglecting the current shunting into the FM layer and the resulting SI-SOI even when $J_{c(FM)} \gg J_{c(NM)}$. $S$ originates from the dampinglike SOT generated by both the SOT from the NM layer and the SI-SOT. Therefore, both $J_{c(NM)}$ and $J_{c(FM)}$ contribute to $S$. In contrast, $A$ is mainly due to the Oersted field generated by $J_{c(NM)}$ only. In case 4A, $J_{c(NM)}$ is much lower than $J_{c(FM)}$ owing to the much lower conductivity of the NM layer. Hence, $\xi_{FMR}$, which is equivalent to $S$ ($\sim J_{s(NM)} + J_{s(FM)}$) divided by $A$ ($\sim J_{c(NM)}$), is significantly overestimated. Figure 4(c) shows the calculated result when $\sigma_{NM}$ is changed (case 4B). The overestimation of $\xi_{FMR}$ becomes pronounced when $\sigma_{NM}$ is low. Further over estimation is expected at the FM/NM interface with spin-splitting states, such as the TI/FM interface. For a TI/FM interface with considerable spin-momentum locking (SML), the electrons change their momentum at the interface, because the Fermi wave number in the topological surface states is considerably smaller than in the FM layer. As a result, SML enhances spin scattering at the interface, which causes enhanced SI-SOT and overestimation of $\xi_{FMR}$. The aforementioned overestimation occurs not only in ST-FMR but also in other methods such as SOT magnetization switching [5], the second-harmonic method [52], and shifting of the hysteresis loop [53], because these methods also calculate the SHA by dividing the signal by $J_{c(NM)}$. Careful attention is strongly needed for a proper understanding of SOT physics in bilayer systems.

## VII. Conclusion

In summary, we observed anomalous sign reversal of the total SOT in Ta(5)/Co($t_{FM}$) bilayers when the Co layer was thicker than 8.5 nm, which was attributed to SI-SOT. Such sign reversal occurs when the SHAs of the NM and FM layers have opposite signs. A theoretical model including the SHE of Co calculated using the spin-diffusion equation well reproduced the experimental results. This model shows that investigating the SHE in various kinds of NM/FM bilayers and choosing an FM layer with the same signs for $\theta_{FM}$ as $\theta_{NM}$ achieves efficient SOT action in an NM/FM bilayer. In addition, we found that neglecting the SI-SOT causes overestimation of $\theta_{NM}$ by two orders for weakly conductive NM layers. The findings in this study enable more reliable SOT estimation, which contributes to development of spin-orbitronics using SOT materials. Furthermore, the significant effect of SI-SOT requires revisiting previous studies claiming high SHAs.

**Acknowledgments**

This work was supported by Japan Society for the Promotion of Science (JSPS) (KAKENHI Grants No. 16H06330, No. 19H02197, No. 20H02607, No. 20K22413, and No. 22H00214), Japan Science and Technology Agency (JST), and Precursory Research for Embryonic Science and Technology (PRESTO) (Grant No. JPMJPR20B2).

**Data availability**

The data that support the findings of this study are available from the corresponding author upon reasonable request.

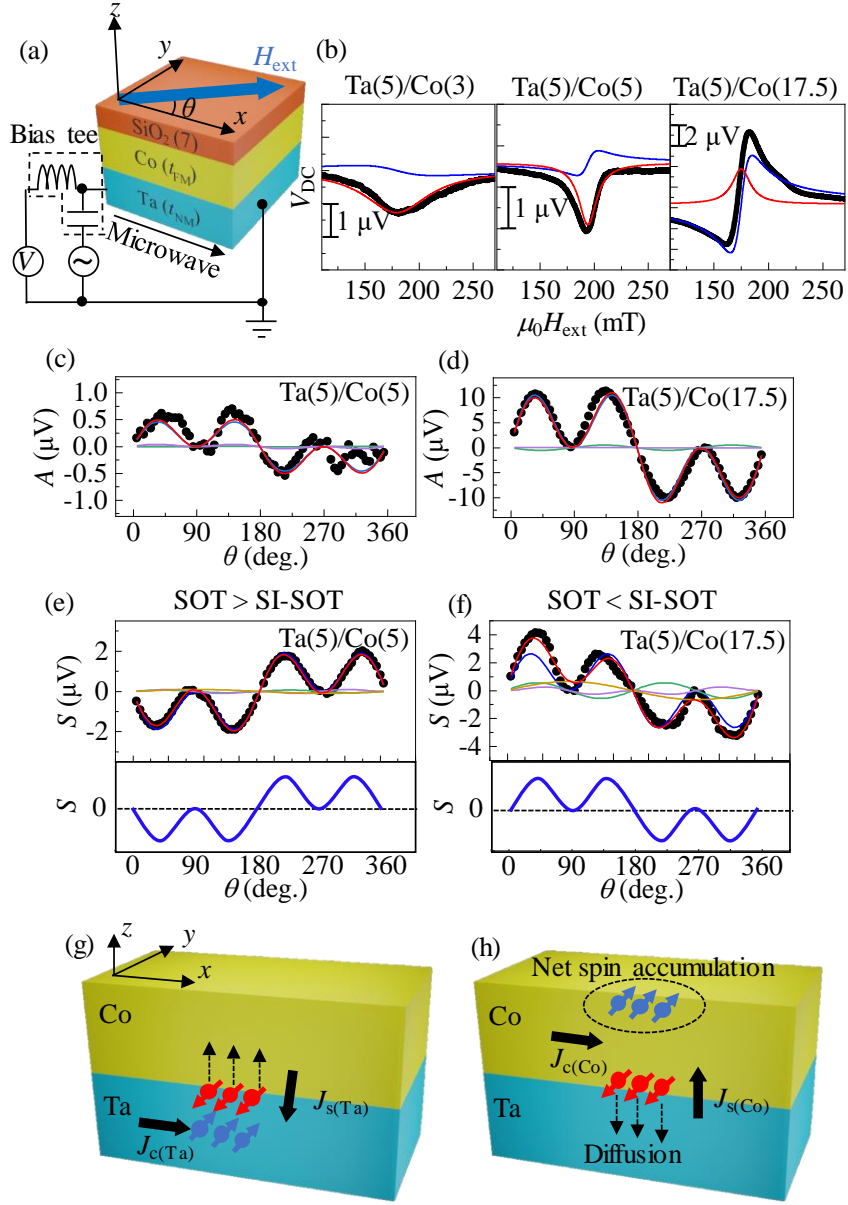

FIG. 1 (a) Schematic of the device structure and the electrical circuit. (b) ST-FMR spectra of (left panel) Ta(5)/Co(3), (middle panel) Ta(5)/Co(5), and (right panel) Ta(5)/Co(17.5) devices. Red and blue curves show the symmetric and the anti-symmetric component obtained by fitting with Eq. (1). (c), (d) Measured $\theta$ dependence of $A$ of (c) Ta(5)/Co(5) and (d) Ta(5)/Co(17.5) devices. (e), (f) (upper panel) Measured and (lower panel) expected $\theta$ dependence of $S$ of (e) Ta(5)/Co(5) and (f) Ta(5)/Co(17.5) devices. Solid lines are the total fitting (red), $\sin2\theta\cos\theta$ (blue), $\sin2\theta$ (green), $\sin2\theta\sin\theta$ (purple), and $\sin\theta$ (brown) terms. (g), (h) Schematics of the spin injection into Co via the SHE in (g) Ta and (h) Co. In the ST-FMR measurements, frequency, $f$, was 13 GHz when $t_{FM}$ = 3 nm and 16 GHz when $t_{FM} \geq$ 5 nm. Signal output power was fixed to 13 dBm.

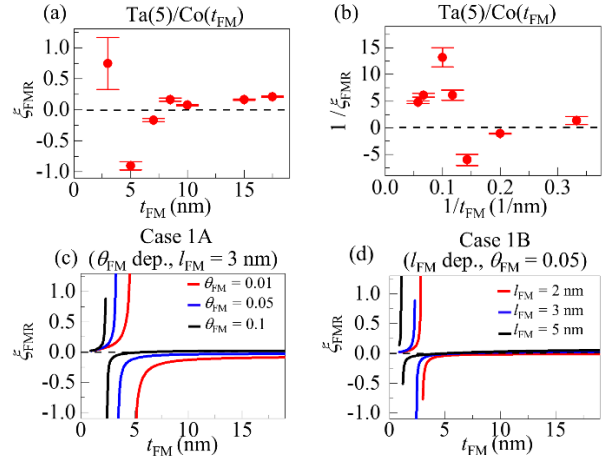

FIG. 2 (a) $\xi_{FMR}$ as a function of $t_{FM}$ and (b) $1/\xi_{FMR}$ as a function of $1/t_{FM}$ for Ta/Co devices. (c) $\theta_{FM}$ and (d) $l_{FM}$ dependences of the $\xi_{FMR}$-$t_{FM}$ curve for case 1.

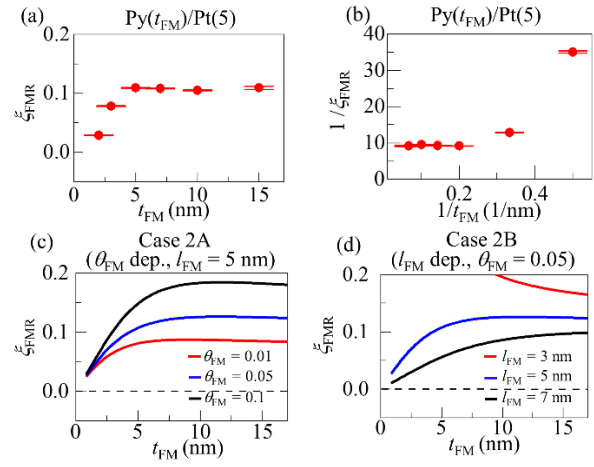

FIG. 3 (a) $\xi_{FMR}$ as a function of $t_{FM}$ and (b) $1/\xi_{FMR}$ as a function of $1/t_{FM}$ for Py/Pt devices. (c) $\theta_{FM}$ and (d) $l_{FM}$ dependences of the $\xi_{FMR}$–$t_{FM}$ curve for case 2. $f$ = 13 GHz, 11 GHz, and 5 GHz when $t_{FM} \geq 10$ nm, 7 nm $\geq t_{FM} \geq 3$ nm, and $t_{FM} = 3$ nm, respectively. Signal output power was fixed to 13 dBm.

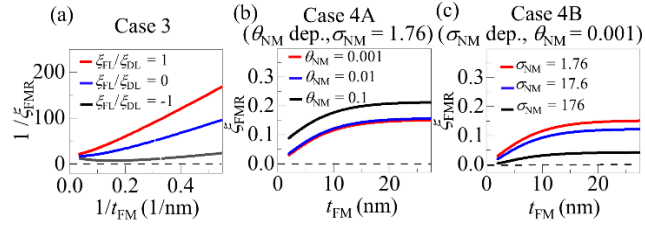

FIG. 4 (a) $\xi_{FL}/\xi_{DL}$ dependence of the $1/\xi_{FMR}$–$1/t_{FM}$ curve for case 3. (b) $\theta_{NM}$ and (c) $\sigma_{NM}$ dependences of the $\xi_{FMR}$–$t_{FM}$ curve for case 4. The parameters for $\xi_{FL}/\xi_{DL}$, $\theta_{NM}$, and $\sigma_{NM}$ used in the calculations are shown in each graph. The parameter units are the same as those of Table I.

Table I. Origin of the angular dependencies of (a)$A$ and (b)$S$. $\sigma_i$ indicates spin-polarization of injected spin current.

(a) $A$ component

| $\theta$ dependence | $\sin2\theta\cos\theta$ | $\sin2\theta$ | $\sin2\theta\sin\theta$ |
|---|---|---|---|
| Torque form | $\mathbf{y}\times\mathbf{M}$ | $\mathbf{M}\times(\mathbf{M}\times\mathbf{z})$ | $\mathbf{x}\times\mathbf{M}$ |
| Origin | FL torque due to $\sigma_y$ | DL torque due to $\sigma_z$ | FL torque due to $\sigma_x$ |
| Oersted field | $\mathbf{y}$ | | $\mathbf{x}$ |
| Undesired effect | | | |

(b) $S$ component

| $\theta$ dependence | $\sin2\theta\cos\theta$ | $\sin2\theta$ | $\sin2\theta\sin\theta$ | $\sin\theta$ |
|---|---|---|---|---|
| Torque form | $\mathbf{M}\times(\mathbf{M}\times\mathbf{y})$ | $\mathbf{z}\times\mathbf{M}$ | $\mathbf{M}\times(\mathbf{M}\times\mathbf{x})$ | |
| Origin | DL torque due to $\sigma_y$ | FL torque due to $\sigma_z$ | DL torque due to $\sigma_x$ | |
| Oersted field | | $\mathbf{z}$ | | |
| Undesired effect | SP-ISHE and ANE | | | SP-ISHE |

Table II. Parameters used in the various cases calculated with Eq. (4).

| Case | $\sigma_{NM}$ ($\times 10^4 (\Omega \cdot m)^{-1}$) | $\sigma_{FM}$ ($\times 10^4 (\Omega \cdot m)^{-1}$) | $l_{NM}$ (nm) | $l_{FM}$ (nm) | $\theta_{NM}$ | $\theta_{FM}$ | $\xi_{FL}/\xi_{DL}$ |
|---|---|---|---|---|---|---|---|
| 1 | 43 | 101 | 1.8 [5] | 2 – 5 | –0.15 [5] | 0.01 – 0.1 | 1 |
| 2 | 176 | 182 | 1.4 [27] | 3 – 7 | 0.32 [27] | 0.01 – 0.1 | –0.2 |
| 3 | 176 | 101 | 3 | 5 | 0.1 | 0.1 | –1,0,1 |
| 4 | 1.76 ~ 176 | 101 | 3 | 5 | 0.001 ~ 0.1 | 0.1 | 0 |

# Anomalous sign inversion of spin-orbit torque in ferromagnetic metal/nonmagnetic bilayer systems due to self-induced spin-orbit torque

Motomi Aoki[1], Ei Shigematsu[1], Ryo Ohshima[1], Teruya Shinjo[1], Masashi Shiraishi[1] and Yuichiro Ando[1,2†]

[1]Department of Electronic Science and Engineering, Kyoto University, Kyoto, Kyoto 615-8510, Japan

[2]PRESTO, Japan Science and Technology Agency, Honcho, Kawaguchi, Saitama 332-0012, Japan

**A. Thickness dependence of the effective magnetization and the damping coefficient**

**B. Contribution of spurious effects**

**C. Shift of the magnetoresistance (MR) curve**

**D. Second harmonic Hall measurements**

**E. Thickness dependence of the MR**

**F. Crystalline structure of Co/Ta bilayers**

**G. ST-FMR measurement for Pt/Fe bilayers**

**H. Spin-orbit torque (SOT) in a single Co layer**

## A. Thickness dependence of the effective magnetization and the damping coefficient

Figure S1(a) shows the effective magnetization, $\mu_0 M_{\text{eff}}$, as a function of the inverse of the thickness of the Co layer, $1/t_{\text{FM}}$, in our Ta(5)/Co($t_{\text{FM}}$) devices. Considering the interfacial magnetic anisotropy, $\mu_0 M_{\text{eff}}$ is expressed as [1],

$$\mu_0 M_{\text{eff}} = \mu_0 M_s - \frac{2K_s}{M_s}\frac{1}{t_{\text{FM}}}, \tag{S1}$$

where $\mu_0$ is the vacuum permeability, $M_s$ is the saturation magnetization, and $K_s$ is the interface perpendicular magnetic anisotropy energy. From the intercept of the linear fit in Fig. S1(a), $\mu_0 M_s$ for Co was estimated to be 1.87 T. Figure S1(b) shows the damping coefficient, $\alpha$, as a function of $1/t_{\text{FM}}$. Two-magnon scattering is very small in a Ta/Co bilayer [2]. In this case, only the spin pumping and/or interfacial spin flipping contribute to the enhancement of $\alpha$, which is expressed as [3]

$$\alpha = \alpha_{\text{int}} + g_{\text{eff}}^{\uparrow\downarrow}\frac{g\mu_B}{4\pi M_s}\frac{1}{t_{\text{FM}}}, \tag{S2}$$

where $\alpha_{\text{int}}$, $g_{\text{eff}}^{\uparrow\downarrow}$, $g$, and $\mu_B$ are the intrinsic damping coefficient, effective mixing conductance, the $g$ factor, and Bohr magneton. From the intercept of the linear fit in Fig. S1(b), $\alpha_{\text{int}}$ in Ta/Co was estimated to be 0.0055. Note that not only spin pumping but also interface scattering and/or spin memory loss contribute to $g_{\text{eff}}^{\uparrow\downarrow}$. To estimate the effective mixing conductance due to spin pumping effect, we measured $\alpha$ for Ta($t_{\text{NM}}$)/Co(5) devices as shown in Fig. S1(c). The $t_{\text{NM}}$ dependence of $\alpha$ is expressed as [4]

$$\alpha = \alpha_{\text{int}} + g_{\text{Ta}}^{\uparrow\downarrow}\frac{g\mu_B}{4\pi M_s}\frac{1}{t_{\text{FM}}}\left(1 - e^{-\frac{2t_{\text{NM}}}{l_{\text{NM}}}}\right), \tag{S3}$$

where $g_{\text{Ta}}^{\uparrow\downarrow}$ is the spin mixing conductance due to the spin pumping into Ta and $l_{\text{NM}} \sim 1.8$ nm [5] is the spin diffusion length of Ta. From fitting using Eq. (S3), $g_{\text{Ta}}^{\uparrow\downarrow}$ was estimated to be $1.7 \times 10^{-19}$ m$^{-2}$.

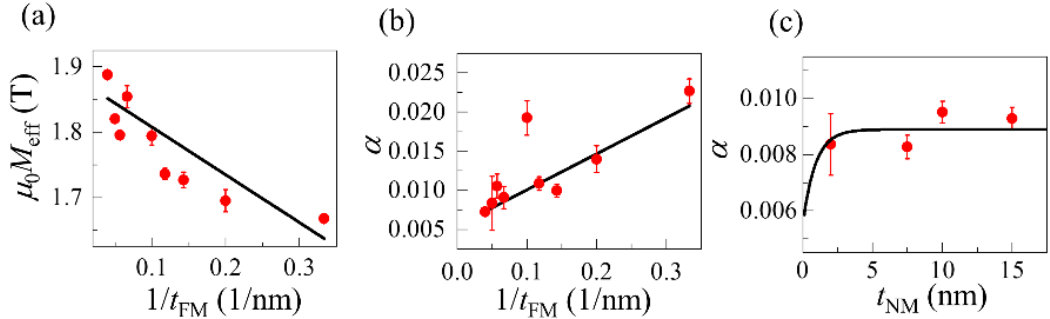

Fig. S1. (a) $\mu_0 M_{\text{eff}}$ and (b) $\alpha$ as functions of $1/t_{\text{FM}}$ in Ta(5)/Co($t_{\text{FM}}$) devices. The black lines show the linear fit. (c) $\alpha$ as a function of $t_{\text{NM}}$ in Ta($t_{\text{NM}}$)/Co(5) devices. The black line shows a fitting curve using Eq. (S3).

## B. Contribution of spurious effects

In the measurement of spin-torque ferromagnetic resonance (ST-FMR), the inverse spin Hall effect induced by spin pumping (SP-ISHE) also generates a symmetric Lorentzian spectrum [6,7]. In our geometry, the sign of the SP-ISHE signal is opposite to that of the symmetric component of the ST-FMR spectrum, $S$. Therefore, SP-ISHE might contribute to the sign inversion of $S$ observed in our Ta/Co devices. The voltage induced by the SP-ISHE, $V_{SP}$, is expressed as [6],

$$V_{SP} = \frac{Le\theta_{NM}g_{Ta}^{\uparrow\downarrow}l_{NM}}{2\pi(t_{NM}\sigma_{NM} + t_{FM}\sigma_{FM})}\tanh(\frac{t_{NM}}{2l_{NM}})\frac{(\tau_{DL}^0)^2\omega_1 + (\tau_{FL}^0 + \tau_{Oe}^0)^2\omega_2}{\alpha^2(\omega^+)^2}, \quad (S4)$$

where $L$, $e$, and $\sigma_{NM(FM)}$ are the channel length, the elementary charge, and conductivity of the NM(FM) layer. $\tau_{DL(FL)}^0 = \xi_{DL(FL)}\mu_B J_{c(NM)}/eM_s t_{FM}$ and $\tau_{Oe}^0 = \gamma\mu_0 J_{c(NM)} t_{FM}/2$, where $\xi_{DL(FL)}$, $J_{c(NM)}$, and $\gamma$ are the dampinglike (DL) [fieldlike (FL)] torque efficiency, current density in the NM layer, and the gyromagnetic ratio, respectively. $\omega_1$, $\omega_2$, and $\omega_+$ are expressed as $\omega_1 = \mu_0\gamma H_{FMR}$, $\omega_2 = \mu_0\gamma(H_{FMR} + M_{eff})$, and $\omega_+ = \omega_1 + \omega_2$ using ferromagnetic the resonance field, $H_{FMR}$. The symmetric component induced by the ST-FMR, $S$, is expressed as [6]

$$S = \frac{I_{rf}R_{AMR}}{2\alpha\omega^+}\tau_{DL}^0, \quad (S5)$$

where $I_{rf}$ is the amplitude of the microwave electric current and $R_{AMR}$ is the amplitude of the anisotropic magnetoresistance (AMR) defined as $R = R_0 + R_{AMR}\sin^2\theta$. Figure S2 shows $|V_{SP}/S|$ as a function of $t_{FM}$ in our Ta(5)/Co($t_{FM}$) devices calculated using Eqs. (S4), (S5), and (3) in the main text, when the self-induced spin orbit torque is not considered, i.e., $\theta_{FM} = 0$. We used the following measured values: Co AMR ratio = 0.5 %, $\sigma_{NM} = 4.3 \times 10^5$ (Ω · m)$^{-1}$, $\sigma_{FM} = 1.0 \times 10^6$ (Ω · m)$^{-1}$. The fixed parameters were $l_{NM} = 1.8$ nm, $l_{FM} = 3$ nm, $\theta_{NM} = -0.15$, and $\xi_{FL}/\xi_{FL} = 1$. At all ranges, $|V_{SP}/S| < 1$, which means the SP-ISHE cannot explain the sign inversion of $S$ observed in our experiments.

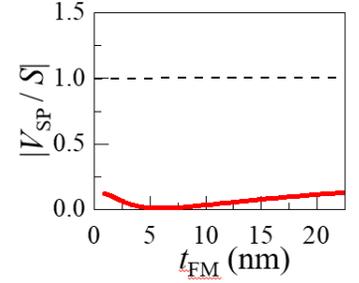

Fig. S2 $|V_{SP}/S|$ as a function of $t_{FM}$ in Ta/Co devices.

## C. Shift of the magnetoresistance (MR) curve

In the main text, we mainly discussed the data obtained via ST-FMR. To ensure that the sign inversion of $S$ originates from the inversion of SOT through another experimental technique, we also detected the sign inversion of SOT from the shift of the MR curve [8]. Figure S3(a) shows the schematics of the measurement. The external magnetic field, $H_{ext}$, was applied along the $+y$ direction in all measurements. Figure S3(b) shows resistance, $R$, as a function of $\mu_0 H_{ext}$ for the Ta(5)/Co(3) sample when $I_{DC}$ = 6 mA (solid line) and –6 mA (dotted line). A positive MR can be observed when the magnetization direction is switched to the opposite direction because of the anisotropic magnetoresistance in Co. Here, we define the switching field, $H_{sw}$, indicated as dots in Fig. S3(b). The magnetization switching takes place by application of effective magnetic field $H_{eff}$, a summation of $H_{ext}$, the Oersted field, $H_{Oe}$ induced by $I_{DC}$, and spin-orbit effective field $H_{SO}$, induced by the FL torque. The spin-orbit effective field by DL torque is along $z$-direction and does not contribute $H_{eff}$. A shift of $H_{sw}$ corresponding to the polarity of $I_{DC}$ was successfully obtained, which is due to non-negligible contribution of $H_{Oe}$ and $H_{SO}$. Figure S3(c) shows $\mu_0 H_{sw}$ as a function of $I_{DC}$. $\mu_0 H_{sw}$ was almost linearly shifted by application of $I_{DC}$. Figure S3(d) shows $\mu_0(H_{sw} - H_{Oe})$ as a function of $I_{DC}$, where $H_{Oe}$ was estimated by calculation. The FL torque efficiency, $\xi_{FL}$, is expressed as [9],

$$\xi_{FL} = \frac{2eM_s t_{FM} t_{NM} w}{\hbar \zeta} \frac{\mu_0 H_{SO}}{I_{DC}}, \quad (S6)$$

where $\zeta = \sigma_{NM} t_{NM} / (\sigma_{NM} t_{NM} + \sigma_{FM} t_{FM})$. The slope of the linear fit in Fig. S3(c) [S3(d)] corresponds to $\mu_0(H_{Oe} + H_{SO})/I_{DC}$ [$\mu_0 H_{SO}/I_{DC}$], which allows us to estimate the thickness dependence of $\xi_{FL+Oe}$ ($\xi_{FL}$) as shown in Fig. S3(e) [S3(f)]. $\xi_{FL}$ changed from negative to positive when $t_{FM} \sim 10$ nm. This sign change is caused by polarity change of the spin current and thus, polarity change of DL torque is also expected at $t_{FM} \sim 10$ nm. This result is consistent with the sign inversion of $\xi_{DL}$ obtained via ST-FMR. Therefore, these results further confirm the existence of the self-induced spin injection due to the spin Hall effect of Co.

We comment on the behavior of antisymmetric component, $A$, of ST-FMR signal. As we discussed in the main text, the sign inversion of $A$ was not taken place at $t_{FM} \sim 10$ nm but at $t_{FM} \sim 5$ nm, which is apparently inconsistent with the MR measurements. In the ST-FMR, $A$ is contributed by both the FL SOT and the torque by $H_{Oe}$. Whereas the FL SOT is proportional to $1/t_{FM}$, $H_{Oe}$ is independent of $t_{FM}$. Therefore, contribution of $H_{Oe}$ is increased by increasing $t_{FM}$. The sign inversion in $A$ is taken place at $t_{FM} \sim 5$ nm when the positive torque of $H_{Oe}$ becomes larger than the negative torque of the FL SOT. Although the polarity of FL SOT is expected to be inverted at $t_{FM} \sim 10$ nm as shown in Fig. S3(f), this cannot be detected in ST-FMR, because total torque for generation of $A$ is already positive because of the dominant contribution of $H_{Oe}$. Therefore, the inversion in $\xi_{FMR}(=S/A)$ was taken place at $t_{FM} \sim 5$ nm for polarity change of $A$ and at $t_{FM} \sim 10$ nm for that of $S$.

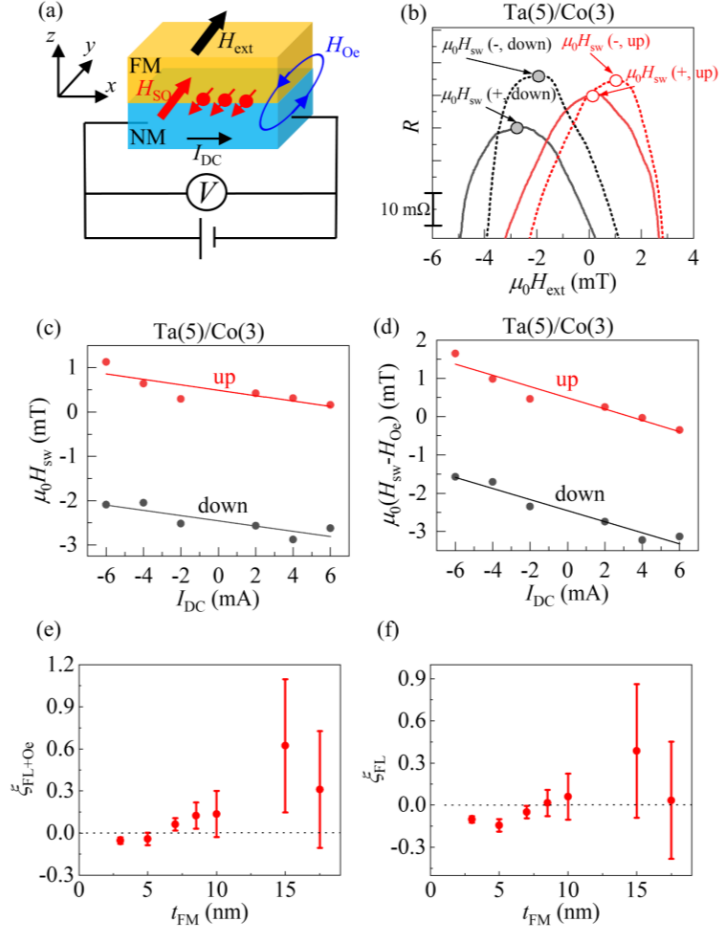

Fig. S3 (a) A schematic of the device structure for measuring the shift of the MR curve. (b) MR curve of Ta(5)/Co(3) when $I_{DC} = \pm 6$ mA. (c) $\mu_0 H_{sw}$ and (d) $\mu_0(H_{sw}-H_{Oe})$ as a function of $I_{DC}$ for Ta(5)/Co(3), where slopes of the fitting are proportional to $H_{SO} + H_{Oe}$ and $H_{SO}$, respectively. (e) $\xi_{FL+Oe}$ and (f) $\xi_{FL}$ as a function of $t_{FM}$ for Ta(5)/Co($t_{FM}$) bilayers.

## D. Second harmonic Hall measurements

We also measured the second harmonic Hall (SHH) signal as another control experiment. Figure S4 shows a schematic of the device structure for the SHH experiment. AC current, $I_{ac}$, with a frequency of 17 Hz was applied using a Keithley 6221 current source. SHH voltage, $V_{xy}^{2\omega}$, was measured using an SR830 lock-in amplifier. When $H_{ext}$ is applied along $\varphi$ direction as shown in Fig. S4, $V_{xy}^{2\omega}$ is expressed as [10,11],

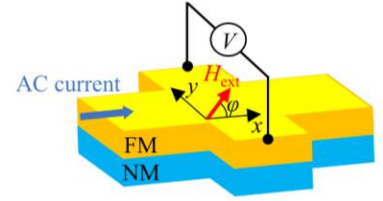

Fig. S4 A schematic of the device structure for SHH experiment.

$$V_{xy}^{2\omega} = V_a \cos\varphi + V_p \cos\varphi \cos 2\varphi. \tag{S7}$$

Here, $V_a$ and $V_p$ are expressed as,

$$V_a = -\frac{I_0 R_{AHE}}{2(H_{ext} + H_k)} H_{DL} + \alpha_{ONE}\mu_0 H_{ext} + V_{ANE}, \tag{S8, a}$$

$$V_p = \frac{I_0 R_{PHE}}{H_{ext}}(H_{SO} + H_{Oe}), \tag{S8, b}$$

where $I_0$ is amplitude of the AC current and $R_{AHE(PHE)}$ is the anomalous (planar) Hall resistance. $\alpha_{ONE}\mu_0 H_{ext}$ and $V_{ANE}$ are the SHH voltage induced by the ordinary Nernst effect and the ANE, respectively. Figures S5(a) [S5(b)] shows $\varphi$ dependence of $V_{xy}^{2\omega}$ for Ta(5)/Co(3) [Ta(5)/Co(10)] at $\mu_0 H_{ext}$ = 1 T. $V_{xy}^{2\omega}$ was successfully fitted using Eq. (S7) as indicated by a red line. Figures S5(c) [S5(d)] shows $\mu_0 H_{ext}$ dependence of $V_a$ for Ta(5)/Co(3) [Ta(5)/Co(10)] with a fitting curve using Eq. (S8, a). Here, we obtained $H_k$ and $R_{AHE}$ from the measurement of the anomalous Hall effect. When $t_{FM}$ = 3 nm, $V_a$ was strongly enhanced in low-magnetic-field region, indicating a large negative $\xi_{DL}$. On the other hand, $t_{FM}$ = 10 nm, $V_a$ was almost linear but slightly suppressed in low-magnetic-field region, indicating a small positive $\xi_{DL}$. Figure S5(e) shows $t_{FM}$ dependence of $\xi_{DL}$ estimated using SHH experiment for Ta/Co devices. Suppression and sign inversion of $\xi_{DL}$ around $t_{FM}$ = 10 nm were successfully observed, which is consistent with the result of both the ST-FMR measurements shown in the main text and the shift of the MR curve shown in section C.

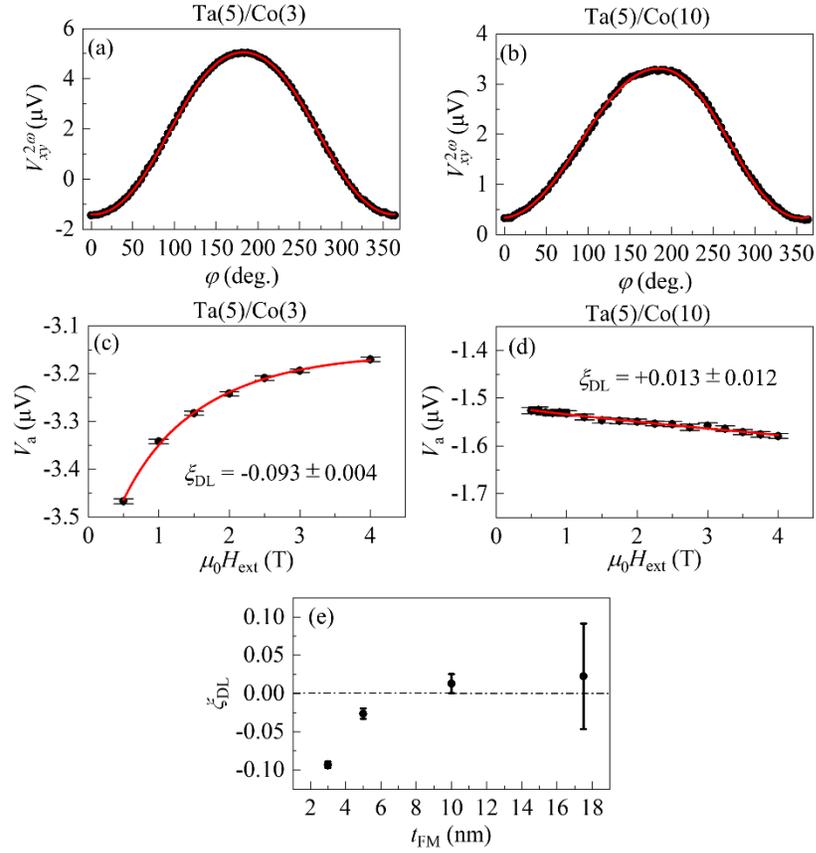

Fig. S5 (a) [(b)] $V_{xy}^{2\omega}$ as a function of $\varphi$ at $\mu_0 H_{ext}$ = 1 T for Ta(5)/Co(3) [Ta(5)/Co(10)] device. A red curve represents the fitting curve using Eq. (S7). (c) [(d)] $V_a$ as a function of $\mu_0 H_{ext}$ for Ta(5)/Co(3) [Ta(5)/Co(10)] device. (e) $t_{FM}$ dependence of $\xi_{DL}$ estimated using SHH measurements. Suppression and sign inversion of $\xi_{DL}$ were observed.

**E. Thickness dependence of the MR**

Since the ST-FMR is the product of the MR effect and current-induced torque, one might expect that sign inversion of $S$ in the ST-FMR is not owing to the self-induced SOT but sign inversion of the MR. Figures S6(a) and S6(b) show MR curve for Ta(5)/Co(3) and Ta(5)/Co(20), respectively, where external magnetic field, $H_{ext}$, was applied perpendicular to the electric current. In both devices, resistance, $R$, exhibited smaller value at large $|\mu_0 H_{ext}|$ because $M$ became perpendicular to the electric current, and larger value at $\mu_0 H_{ext} \sim 0$ mT because magnetic domain parallel to the electric current was created. This indicates that polarity of MR effect does not change between these two devices. MR ratio, $\Delta R/R$, as a function of $t_{FM}$ is shown in Fig. S6(c), where $\Delta R$ is the difference between maximum ($\mu_0 H_{ext} \sim 0$ mT) and minimum (large $|\mu_0 H_{ext}|$) values as shown in Fig. S6(a) and $R$ is average value of the resistance. $\Delta R/R$ was always positive irrespective of $t_{FM}$, indicating that sign inversion of $S$ in the ST-FMR does not originate from sign inversion of the MR. We note that even if sign and/or magnitude of the MR changes, it does not affect $\xi_{FMR}$, because $\xi_{FMR}$ is obtained from the ratio of $S$ and $A$, both of which are proportional to the MR ratio.

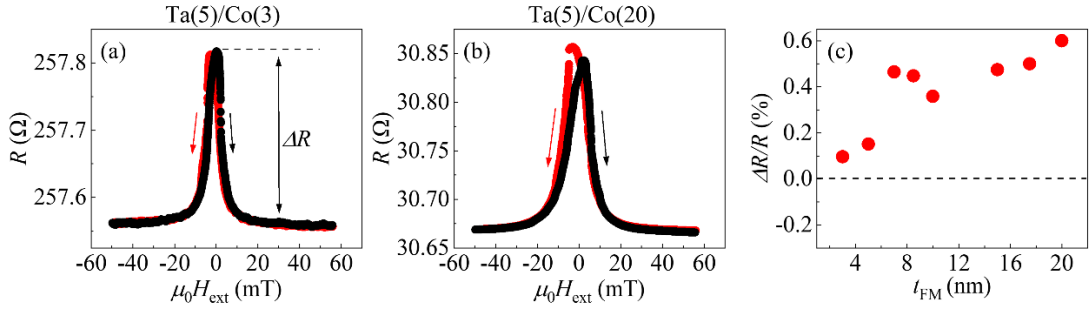

Fig. S6 $R$ as a function of $\mu_0 H_{ext}$ for (a) Ta(5)/Co(3) and (b) Ta(5)/Co(20), where $I_{DC}$ = 2 mA and 8 mA, respectively. (c) $\Delta R/R$ as a function of $t_{FM}$.

**F. Crystalline structure of Co/Ta bilayers**

Figures S7(a) and S7(b) show X-ray diffraction (XRD) patterns of Ta (5 nm)/Co(5 nm) and Ta (5 nm)/Co (17.5 nm), respectively. No peak other than that originating from MgO substrate was observed indicating our Ta/Co films are amorphous. Figure S7(c) shows comparison of XRD patterns of Ta (5 nm)/Co (5 nm) and Ta (5 nm)/Co (17.5 nm). No visible difference between Fig. S7(a)[S7(b)] and S7(c)[S7(d)] was observed. Therefore, crystal structure of Co layer for $t_{FM}$ = 5 and 17.5 nm were almost identical.

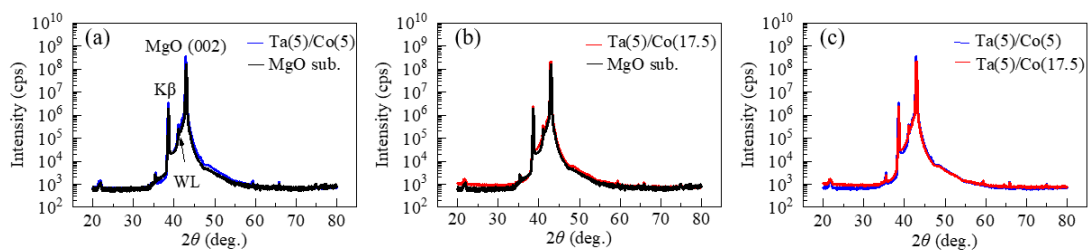

Fig. S7 (a)[(b)] XRD spectrum of Ta(5)/Co(5) [Ta(5)/Co(17.5)]. (c) Comparison of Ta (5)/Co(5) and Ta (5 nm)/Co (17.5 nm).

## G. ST-FMR measurement for Pt/Fe bilayers

We also measured ST-FMR for Pt/Fe bilayer. Because sign of the SHA in Pt and Fe are known to be opposite to each other [12,13], cancellation of the SOT by the SI-SOT and resulting suppression of $\xi_{FMR}$ is expected at large $t_{FM}$. Figures S8(a) and S8(b) show $\theta$ dependence of $A$ and $S$ of Pt (15 nm)/Fe (3 nm) obtained in the same measurement setup as that of Ta/Co bilayers. Solid lines indicate $\sin2\theta\cos\theta$ (red), $\sin2\theta$ (green), $\sin2\theta\sin\theta$ (purple), and $\sin\theta$ (brown) terms. By taking the ratio between $\sin2\theta\cos\theta$ terms in $A$ and $S$, $\xi_{FMR}$ was calculated to be $0.132\pm0.001$. Figures S8(c) and S8(d) show $\theta$ dependence of $A$ and $S$ for Pt (15 nm)/Fe (12 nm). A red dotted line in Fig. S8(d) shows $\sin2\theta\cos\theta$ term enlarged by an order of magnitude. $\xi_{FMR}$ was strongly suppressed and shows opposite sign ($-0.009\pm0.005$), which is the typical behavior of the NM/FM bilayers with opposite SHAs as discussed in the main text.

Fig. S8 $\theta$ dependence of (a)[(c)] $A$ and (b)[(d)] $S$ for Pt(15)/Fe(3) [Pt(15)/Fe(12)]. A red dotted line indicates $\sin2\theta\cos\theta$ term enlarged by an order of magnitude.

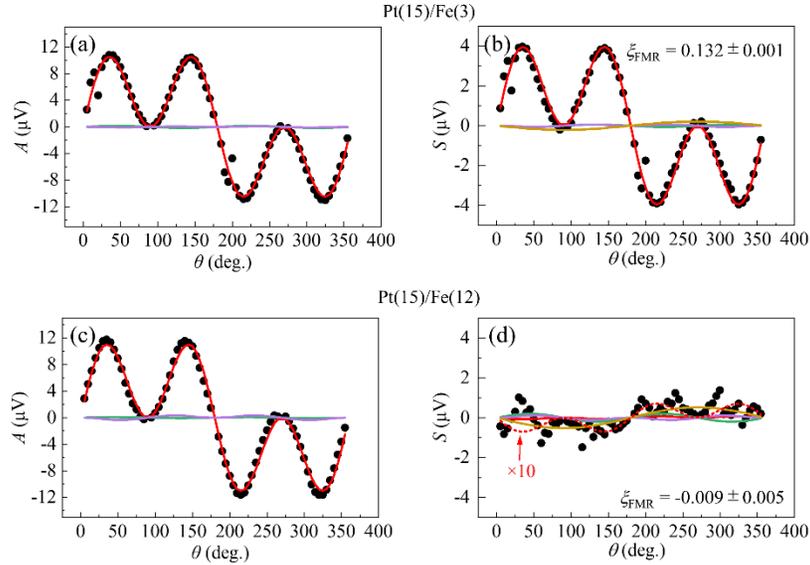

**H. SOT in a single Co layer**

According to previous researches, non-zero SOT is observed in a single FM layer without any adjacent layer if inversion symmetry is broken by interface and/or strain [13–15]. Recent paper revealed that such single-layer SOT becomes salient in a NM/FM bilayer system when the conductivity of the NM layer is low [16]. To exclude the contribution from single-layer SOT, we measured the ST-FMR spectrum for Co(10 nm)/SiO$_2$(7 nm) layer. As a result, no resonance spectrum was observed at any $\theta$, indicating single-layer SOT in the Co layer is negligibly small as shown in Figure S9. Our result indicates that contribution from SI-SOT in NM/FM bilayer systems is significant even when the SOT of single FM layer is negligibly small because of spin-current absorption into an NM layer with large spin-orbit coupling such as Ta.

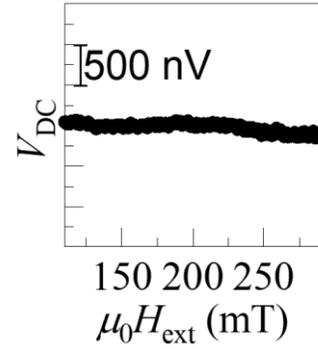

Fig. S9 ST-FMR signal of Co(10 nm)/SiO$_2$(7 nm) film at $\theta = 45°$ and with 16 GHz microwave.